\documentclass[aps,prl,preprint,groupedaddress]{revtex4}

\usepackage{graphicx}

\begin{document}


\title{Synthesis of as-grown superconducting MgB$_2$ thin films by molecular beam epitaxy in UHV conditions}


\author{Y. Harada, M. Udsuka, Y.Nakanishi, and M. Yoshizawa}
\email{yharada@isop.ne.jp}
\affiliation{Department of Materials Science and Engineering, 
        Iwate University, \\ 
        Ueda 4-3-5, Morioka, 020-8551, Japan}


\date{\today}

\begin{abstract}
As-grown superconducting MgB$_2$ thin films have been grown on SrTiO$_3$(001), MgO(001), and Al$_2$O$_3$(0001) substrates by a molecular beam epitaxy (MBE) method with novel co-evaporation conditions of low deposition rate in ultra-high vacuum. 
The structural and physical properties of the films were studied by RHEED, XRD, electrical resistivity measurements, and SQUID magnetometer. 
The RHEED patterns indicate three-dimensional growth for MgB$_2$. 
The highest {\it T}$_c$ determined by resistivity measurement was about 36K in these samples. 
And a clear Meissner effect below {\it T}$_c$ was observed using magnetic susceptibility measurement. 
We will discuss the influence of B buffer layer on the structural and physical properties
\end{abstract}


\maketitle


\section{Introduction}
The discovery of superconductivity at 39K in magnesium diboride (MgB$_2$) has attracted great interest in science and technology, since it shows the highest transition temperature ({\it T}$_c$) among intermetallic compound superconductors \cite{aki1}. 
There have already been several reports on the preparation technique of MgB$_2$ thin film so far \cite{kang1, Eom1}. 
Many of them require a post-annealing process to improve physical properties and attain superconductivity. 
As-grown process is desired for the fabrication of tunnelling junctions and multilayers for MgB$_2$. 
Several groups have reported as-grown superconducting MgB$_2$ films so far \cite{ueda1, Jo1, erven1, shimakage1}. 
In as-grown method, low temperature synthesis, which is needed to depress high volatility of Mg makes it difficult to produce MgB$_2$ thin films with high crystallinity. 
For this reason, {\it T}$_c$ of thin MgB$_2$ film grown by molecular beam epitaxy (MBE) is limited to be around 35K at best. 
On the contrary, chemical vapor deposition (CVD) method, which enables a high-temperature growth process of the film, succeeds in making the film with high {\it T}$_c$ = 39K \cite{zeng1}. 

It is considered that MgB$_2$ should be grown up in vacuum as high as possible to get good crystallinity, because the quality of MgB$_2$ thin films is considerably affected by residual gases \cite{ueda2}. 
It is a promising way for device application. 
Therefore, it is desired to establish the low rate deposition technique at low growth temperature to produce high quality films.

In this paper, we will report the growth of MgB$_2$ thin film using MBE. 
The {\it T}$_c$ of the obtained superconducting MgB$_2$ thin films is 36K, which is comparable to other reports by MBE method. 
We will also discuss the influence of B buffer layer on the structural and physical properties.

\section{Experimental}
The MgB$_2$ films were grown on SrTiO$_3$(001) (STO), MgO(001), and Al$_2$O$_3$(0001) substrates in an MBE chamber with a base pressure of 4$\times 10$$^{-10}$ Torr. 
Mg and B metals were used as the evaporation sources. 
The both purities of Mg and B were 3N. 
Mg was evaporated from a Knudsen cell, and B by electron-beam. 
The deposition rate was controlled by a quartz-crystal monitor (QCM) with the flux ratio of Mg to B varied from 2 to 10 times as high as the nominal flux ratio to compensate Mg loss, and the deposition rate of B was fixed at 0.5 \AA/s. 
The thickness of films was typically 1000\AA. 
The substrate temperature ({\it T}$_s$) was varied between 250 and 280$^\circ$C. 
The background pressure during the growth was better than 4$\times 10$$^{-9}$ Torr. 
We chose the temperature by the work of Liu et al. \cite{liu1}. 
All of the films were covered by 50\AA-thick Mg cap layer to avoid oxidation. 
The crystal structure characterized by in-situ reflected high energy electron diffraction (RHEED) and ex-situ X-ray diffraction (XRD: 2$\theta$-$\theta$scan). 
The resistivity and magnetic susceptibility measurements were carried out by the standard four-probe technique and SQUID magnetometer respectively.

\section{Results and Discussion}
The films prepared at above 280$^\circ$th an Mg/B ratio of 4 were significantly Mg deficient, and showed semiconducting properties with no trace of superconductivity. 
Superconductivity was achieved only at {\it T}$_s$ = 250$^circ$C and with an Mg/B ratio of 8-10. 
Figure 1 shows a typical temperature dependence of electrical resistivity for MgB$_2$ thin films deposited on the various substrates, in which Mg/B ratio was 9. 
The values of {\it T}$_{c,onset}$-{\it T}$_{c,zero}$ were 35.0-34.0, 33.5-31.2, and 32.3-28.2K for the films on STO (001), MgO (001), and Al$_2$O$_3$ (0001), respectively. 
Here, the {\it T}$_{c,onset}$ was defined as the temperature at which resistance abruptly started to drop, and the {\it T}$_{c,zero}$ was defined as the temperature at which resistivity reached zero. 
The normal state resistivity at 300K was 263 (STO), 163 (MgO), and 14.5 $\mu\Omega$cm (A$_2$O$_3$), respectively. 
The residual resistivity ratio ({\it RRR}) was 1.15 (STO), 1.14 (MgO), and 1.21 (Al$_2$O$_3$), respectively. 
Here, {\it RRR} was defined as the ratio of resistivities at 300 and 40K. 
The {\it RRR} of our films were similar to those of other reported films. 
It should be noted that the resistivity of the films grown on Al$_2$O$_3$ is lower than that on STO and MgO.

The structure and crystallinity were characterized by RHEED and XRD. 
Figure 2 shows the XRD pattern of MgB$_2$ thin film on STO substrate. 
The inset is the RHEED pattern of this film.  
XRD results show no characteristic peak of the film except for substrate peaks. 
The dotted line shows the expected peak position of MgB$_2$(002). 
The RHEED shows spotty and ring pattern, suggesting the three-dimensional growth. 
These results indicate poor crystallinity of the films. 
Ueda and Naito reported that even the films with poor crystallinity exhibit superconductivity \cite{ueda1}.

Figure 3 shows the zero-field-cooled (ZFC) and the field-cooled (FC) dc magnetization ({\it M}-{\it T}) curves of MgB$_2$ thin film on STO substrate in a 10-Oe field applied normal to the film plane. 
The size of the sample was 3mm$^2$. 
A clear Meissner effect was observed around 33K. 
The ZFC curve shows the broader diamagnetic transition compared to that of the resistivity data.

Next, we examined the effect of B buffer layer. 
Hur et al. reported that boron buffers brought about an increase of the transition temperature up to 41.7K and acted as a precursor of MgB$_2$ \cite{hur1}. 
For these reasons, we inserted boron buffer layer to improve the critical temperature and the crystallinity of the films. 
Here, the buffer layer was fixed to be 250\AA thick.

Figure 4 shows the temperature dependence of resistivity of superconducting MgB$_2$ thin films on different substrates with boron buffer layers. 
The deposition rates of Mg and B were fixed at the same value, previously reported by our groups. 
The {\it T}$_c$ was slightly enhanced by the insertion of boron buffer in the case of STO substrate. 
Same investigation was also performed on the other substrates. 
The {\it T}$_{c,zero}$ of the MgB$_2$ films on MgO(001) and Al$_2$O$_3$(0001) substrates with insertion of B buffer layer are 29.0 and 29.4K, respectively. 
The {\it T}$_c$ was slightly decreased with insertion of B buffer layer in the case of MgO substrate, whereas {\it T}$_c$ was almost unchanged on Al$_2$O$_3$ substrate. 
The {\it T}$_c$ of these films is lower than that of the bulk form.

Now we discuss the difference of structural and physical properties by insertion of B buffer. 
In RHEED patterns of B buffer layer, halo pattern was observed, and MgB$_2$ showed spot-like pattern. 
In XRD measurements, no peak of MgB$_2$ was observed. 
These results suggested that B buffer layers were amorphous or polycrystalline forms. 
Therefore, the insertion of B buffer layer did not improve the crystallinity of the MgB$_2$ thin films. 
Transition temperature, transition width ($\Delta${\it T}$_c$), normal state resistivity at 300K and {\it RRR} of our films are summarized in Table I. 
An increase of {\it T}$_c$ by insertion of B buffer layer was not observed unlike Hur et al.'s report \cite{hur1}. 
The resistivity at 300K was remarkably decreased by the insertion of B buffer layer for MgO substrate. 
For STO substrate, the resistivity slightly decreased with B buffer layer. 
Probably it is ascribed to the reduction of grain boundary, that is, enlargement of crystal grain size. 
The resistivity was showed about 4 times increase by the insertion of B buffer layer for Al$_2$O$_3$ substrate. 
Other physical properties of MgB$_2$ such as {\it RRR} and $\Delta${\it T}$_c$ were not affected by the insertion of the amorphous B buffer layers.

\section{Summary}
We succeeded in making the as-grown superconducting MgB$_2$ thin films on SrTiO$_3$ (001), MgO (001), and Al$_2$O$_3$ (0001) substrates in the condition of slow deposition rate and ultra-high vacuum for the first time. 
The RHEED patterns indicated that MgB$_2$ films were grown in three-dimensional way. 
The highest {\it T}$_{c,onset}$ with sharp transition width of the films was 36K for STO substrate. 
Furthermore, we investigated the effect of insertion B buffer layer. 
The B buffer layers were amorphous. 
The transport and magnetization measurements indicated that the insertion amorphous B buffer layer did affect the resistivity at 300K, and did not affect the physical properties such as {\it T}$_c$, {\it  RRR}, and $\Delta${\it T}$_c$.


\bibliography{mgb2}

\newpage



%

\begin{table}[tp]
\begin{center}
\label{table1}\caption[super]{Transition temperature, transition temperature width, normal state resistivity at 300K, and residual resistivity ratio ({\it RRR}) of the MgB$_2$ films deposited in this work.}
\smallskip
\begin{tabular}{ccccccc}\hline
Substrate & Buffer & {\it T}$_{c.onset}$ & {\it T}$_{c,offset}$ & $\Delta${\it T}$_c$ & $\rho$(300K) & {\it RRR} \\ \hline
STO & & 35.0 & 34.0 & 1.0 & 263 & 1.15 \\ \cline{1-1} \cline{3-7}
MgO & N/A & 33.5 & 31.2 & 2.3 & 162 & 1.14 \\ \cline{1-1} \cline{3-7}
Al$_2$O$_3$ & & 32.3 & 28.2 & 4.1 & 14.5 & 1.21 \\ \hline
STO & & 36.2 & 34.5 & 1.5 & 209 & 1.19 \\ \cline{1-1} \cline{3-7}
MgO & B(250\AA) & 32.5 & 29.0 & 3.0 & 74.0 & 1.10 \\ \cline{1-1} \cline{3-7}
Al$_2$O$_3$ & & 32.3 & 29.4 & 2.9 & 63.0 & 1.24 \\ \hline
\end{tabular}
\end{center}
\end{table}

\newpage

\begin{figure}[htp]
	\includegraphics[width=0.8\linewidth]{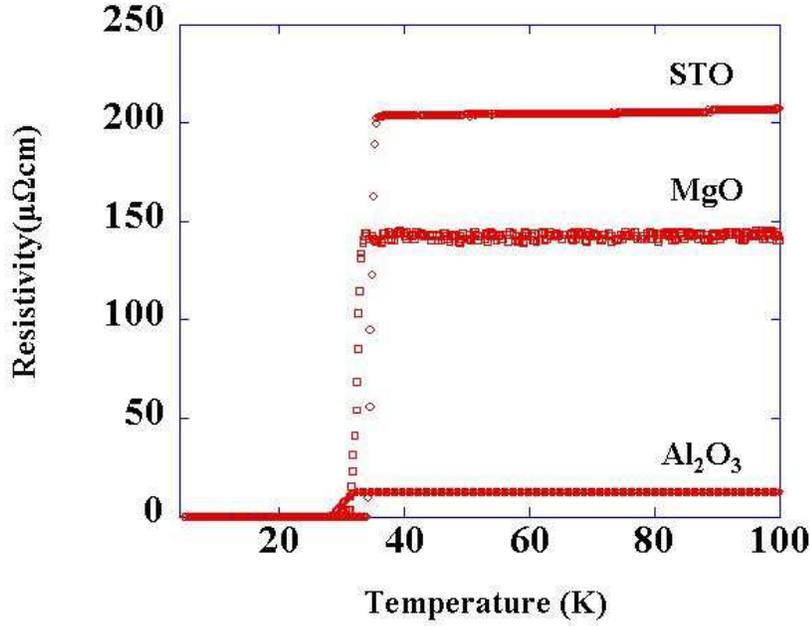}
	\caption{Temperature dependence of resistivity of MgB$_2$ thin films on MgO(001), SrTiO$_3$(001), and Al$_2$O$_3$(0001) substrates. These films were made at 250$^\circ$C with Mg deposition rate of 4.5\AA/sec and B deposition rate of 0.5\AA/sec.}
\end{figure}
\begin{figure}[tp]
	\includegraphics[width=0.8\linewidth]{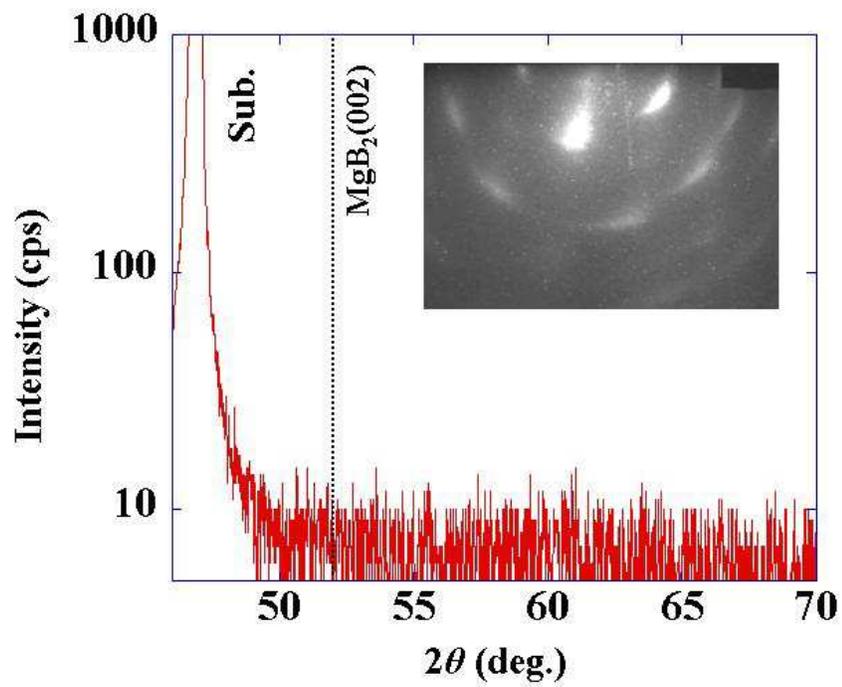}
	\caption{XRD pattern of MgB$_2$ thin film on SrTiO$_3$(001) substrate. The inset is a RHEED pattern of this film.}
\end{figure}
\begin{figure}[tp]
	\includegraphics[width=0.8\linewidth]{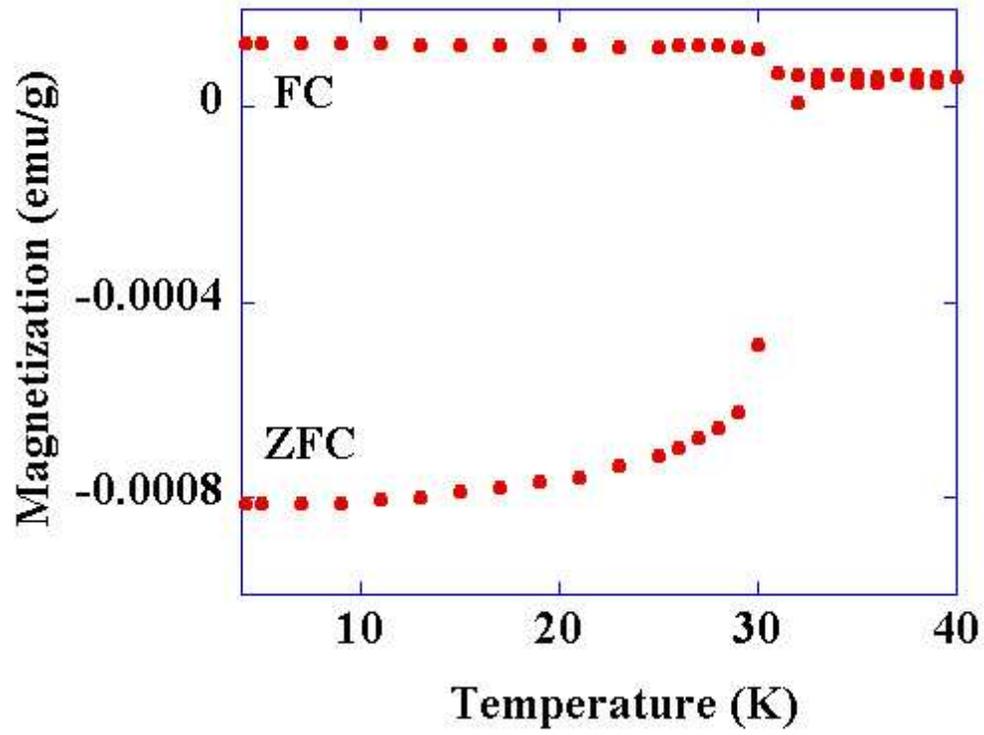}
	\caption{Zero-field-cooled (ZFC) and field-cooled (FC) dc magnetization curves of MgB$_2$ thin film on SrTiO$_3$(001) substrate in the field of 10 Oe applied perpendicular to the film plane.}
\end{figure}
\begin{figure}[tp]
	\includegraphics[width=0.8\linewidth]{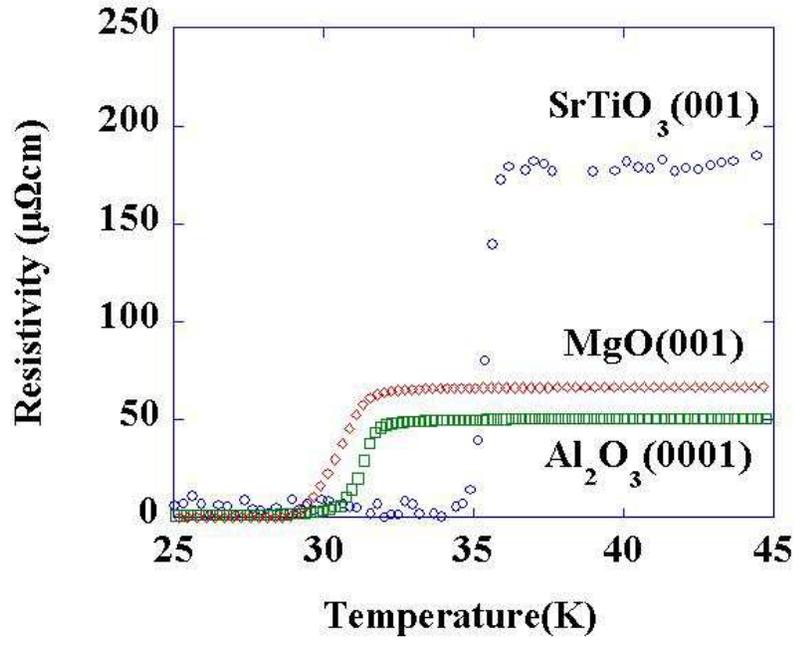}
	\caption{Temperature dependence of resistivity for superconducting MgB$_2$ thin films deposited on different substrates with boron buffer layers.}
\end{figure}





\end{document}